# Swimming strategy of settling elongated microswimmers by reinforcement learning

Jingran Qiu, Weixi Huang, Chunxiao Xu, and, Lihao Zhao*

*AML, Department of Engineering Mechanics, Tsinghua University, 100084 Beijing, China*

**Abstract.** Particular types of plankton in aquatic ecosystems can coordinate their motion depending on the local flow environment to reach regions conducive to their growth or reproduction. Investigating their swimming strategies with regard to the local environment is important to obtain in-depth understanding of their behavior in the aquatic environment. In the present research, to examine an impact of the shape and gravity on a swimming strategy, plankton is considered as settling swimming particles of ellipsoidal shape. The Q-learning approach is adopted to obtain swimming strategies for *smart* particles with a goal of efficiently moving upwards in a two-dimensional steady flow. Strategies obtained from reinforcement learning are compared to those of *naive* gyrotactic particles that is modeled considering the behavior of realistic plankton. It is found that elongation of particles improves the performance of upward swimming by facilitating particles' resistance to the perturbation of vortex. In the case when the settling velocity is included, the strategy obtained by reinforcement learning has similar performance to that of the naive gyrotactic one, and they both align swimmers in upward direction. The similarity between the strategy obtained from machine learning and the biological gyrotactic strategy indicates the relationship between the aspherical shape and settling effect of realistic plankton and their gyrotactic feature.

## I. Introduction

Various plankton are ubiquitous in the marine environment. Many of them are able to swim and coordinate their motion by using special cellular structures, such as flagella [1, 2]. The impact of the swimming motion on the distribution and deposition of plankton in the flow has been considered as an important question. Active plankton are often considered as swimming particles, which can rotate themselves toward a particular direction and swim along it accordingly [3, 4]. Most of the previous studies have been focusing on a naive particle model, in which the particles are supposed to be motile, but are unable to sense or react with the ambient flow environment. Gyrotactic particles, for example, adjust their alignment by a gravitational torque induced by the bias of centers of mass and hydrodynamic force [3]. The motion and deposition of gyrotactic particles have been investigated both in laminar [5-9] and turbulent flows [10-13]. These results show that the motion and clustering patterns of particles are altered by the swimming velocity and gyrotaxis. However, in reality, plankton are able to obtain ambient information, such as temperature [14, 15], light intensity [16, 17], and also motion and acceleration of the ambient flow [18-20]. Based on this information, plankton can coordinate their motion to reach locations that are conducive to their growth and reproduction. Investigating different strategies of how plankton react with the flow environment is of great importance to the advanced understanding of their habits and the physics of the particle-fluid interaction.

* Corresponding author (email: zhaolihao@mail.tsinghua.edu.cn)

Recently, the machine learning approach has been widely utilized to investigate the behavioral strategies of active matters in fluids. Novati *et al.* [21] studied the optimal swimming strategy for a fish to take advantage of wakes; Gazzola *et al.* [22] presented investigation of the fish schooling strategy aiming to minimize the energy consumption of individuals. Similar research on birds or gliders soaring in turbulence was performed by Reddy *et al.* [23]. With regard to active micro-swimmers, studies were recently conducted on the swimming strategy for smart active particles, which can obtain the local flow information and adjust their motion to achieve predefined goals [24-26]. These studies demonstrate the great potential of the reinforcement learning (RL) approach in terms of searching effective swimming strategies for tiny active particles. The pioneer research work presented by Colabrese *et al.* [24] considers point-like, inertia-less spheres in a two-dimensional stationary Taylor-Green vortex (TGV) flow. Smart particles can obtain the information of local flow vorticity, as well as instantaneous particle orientation, and swim with a constant velocity in a direction that is actively determined by the particle itself. They tested particles with different swimming velocity and stability of orientation, and found that smart particles learn to swim upwards much better than naive gyrotactic ones. The success of training micro-swimmers was soon extended to three-dimensional flows [26]. Satisfying outcome was reached that smart particles learn to avoid vortexes trapping and find proper regions where they can take advantage of the flow and swim upwards faster. Colabrese *et al.* [25] recently applied RL on inertial spherical particles, which can actively control their density by changing the particle volume. This kind of particles successfully learn to sample flow regions with specific vorticity, which is artificially given in advance.

These three studies confirm the feasibility of the RL algorithm even in the case of relatively complex configurations, such as time-dependent or three-dimensional flows. However, only the ideal particle models are considered in these studies. Shape and gravity effects are neglected in these studies. In nature, however, both effects are significant for plankton. First, plankton have a variety of shapes that have a considerable influence on the particle behavior in both laminar [7] and turbulent flows [27]. Second, the settling effect caused by gravity is considered to be important in the vertical migration of plankton [28]. Both effects might be important to realistic plankton, and, thereby, are correlated to their swimming strategy. Therefore, driven by the curiosity about the behavior of marine plankton, we aim to deepen our understanding of how the aspherical shape and gravity settling will affect the swimming strategy of a microscopic swimming particle.

To address the aforementioned question, we focus on further development of the concepts proposed in the previous work [24] and investigate the shape and gravity effects on swimming strategies for elongated, inertia-less swimming particles in a two-dimensional stationary TGV flow. Vertical migration is a common but important phenomenon of marine plankton [29]. Therefore, the goal of particles is to swim upwards as fast as possible. We employ RL to obtain approximately optimal swimming strategies, analyze the shape and gravity effects on the characteristics of learned strategies, and compare the learned strategies with the naive gyrotactic ones. In Section 2, we introduce the governing equations of microswimmers and describe the RL algorithm. In Section 3, we analyze the shape and gravity effects on swimming strategies. Finally, we draw the conclusions in Section 4.

**Table 1** Computation parameters.

| Parameter | Symbol | Value |
|---|---|---|
| Length scale of TGV flow | $L_0$ | $5\times10^{-4}$ (m) |
| Velocity scale of TGV flow | $u_0$ | $2\times10^{-3}$ (m/s) |
| Particle-to-fluid density ratio | $\rho_p/\rho_f$ | 1.04 |
| Kinematic viscosity of the fluid | $\nu$ | $1\times10^{-6}$ (m$^2$/s) |
| Gravitational acceleration | $g$ | 9.8 (m/s$^2$) |
| Translational diffusion coefficient | $D_t$ | $1\times10^{-9}$ (m$^2$/s) |
| Rotational diffusion coefficient | $D_r$ | $4\times10^{-4}$ (s$^{-1}$) |

**Table 2** Case settings for particles of different shapes.

| $\lambda$ | $a$ (μm) | $B$ (s) | $v_{swim}$ (μm/s) | $v_s(0)$ (μm/s) | $v_s(\pi/2)$ (μm/s) |
|---|---|---|---|---|---|
| 1.0 | 50.00 | 2.5 | 600.0 | 217.8 | 217.8 |
| 1.5 | 44.00 | 2.5 | 600.0 | 229.6 | 211.8 |
| 2.0 | 40.55 | 2.5 | 600.0 | 237.9 | 207.7 |
| 3.0 | 36.57 | 2.5 | 600.0 | 248.8 | 202.3 |
| 5.0 | 32.69 | 2.5 | 600.0 | 260.8 | 196.3 |
| 10.0 | 28.83 | 2.5 | 600.0 | 273.5 | 189.9 |
| 20.0 | 26.02 | 2.5 | 600.0 | 282.7 | 185.3 |

## II. Methodology

In the present work, we consider particles swimming in a two-dimensional, steady, and spatially periodic TGV flow. The TGV flow is defined as follows:

$$u_x = \frac{u_0}{2}\cos\frac{x}{L_0}\sin\frac{y}{L_0}, \tag{1}$$

$$u_y = -\frac{u_0}{2}\sin\frac{x}{L_0}\cos\frac{y}{L_0}, \tag{2}$$

$$\omega_z = -\frac{u_0}{L_0}\cos\frac{x}{L_0}\cos\frac{y}{L_0}, \tag{3}$$

where $u_x$, $u_y$, and $\omega_z$ are the flow velocity in $x$ and $y$ directions and the vorticity, respectively. In addition, $u_0$ and $L_0$ are the characteristic velocity and length of the flow, which are adopted from the Kolmogorov scales of turbulence in the ocean [30] (at the dissipation rate of $1.6\times10^{-5}$ m$^2$s$^{-3}$) to provide a similar velocity gradient to ocean turbulence. Microswimmers are considered as ellipsoidal, settling particles, whose length scale is small relative to that of the flow, and the density is close to the fluid. Parameters of marine plankton vary in a large range. Therefore, we consider only the typical values of their swimming and settling velocity, size, and density [20, 28, 31, 32]. Parameters for computation and parameters of particles of different shapes are presented in Tables 1 and 2, respectively. Based on these parameters, the particle is in low-Reynolds number regime, and the inertia-less point-particle model is justified [24]. The settling velocity of particles is comparable relative to the swimming velocity. Therefore, we introduce the Stokes settling velocity to account for the gravity effect. The particle motion is governed by the following equations [24, 33]:

$$\dot{\boldsymbol{p}} = \frac{1}{2B}\left[\boldsymbol{k} - (\boldsymbol{k}\cdot\boldsymbol{p})\boldsymbol{p}\right] + \frac{1}{2}\boldsymbol{\omega}\times\boldsymbol{p} + \frac{\lambda^2-1}{\lambda^2+1}\left[\boldsymbol{I} - \boldsymbol{pp}\right]\cdot\boldsymbol{S}\cdot\boldsymbol{p} + \boldsymbol{d}_{\mathrm{r}}, \tag{4}$$

$$\dot{\boldsymbol{x}} = \boldsymbol{v} + \boldsymbol{d}_{\mathrm{t}} = \boldsymbol{u} + v_{\mathrm{swim}}\boldsymbol{p} + \boldsymbol{v}_{\mathrm{s}} + \boldsymbol{d}_{\mathrm{t}}\,. \tag{5}$$

Here, $\boldsymbol{p}$ denotes the unit direction vector along the particle symmetric axis in the inertial frame. The first term on the right hand side of Eq.(4) stands for the effect of the particle active alignment. $B$ denotes the characteristic timescale of particle rotating to align with the direction vector of active alignment $\boldsymbol{k}$. We consider two kinds of particles. *Naive gyrotactic particles* cannot react with the local flow environment, and thereby, $\boldsymbol{k}$ is constantly opposite to the gravity direction. In turn, according to certain strategies, *Smart particles* can actively adjust $\boldsymbol{k}$ based on their instantaneous orientation and vorticity of the local flow. The second and third terms denote the contributions of the fluid vorticity and deformation rate, respectively. Here, $\boldsymbol{\omega}$ and $\boldsymbol{S}$ are the fluid vorticity and the deformation rate tensor at the particle position, respectively. The aspect ratio, $\lambda$, is defined as the ratio of particle's major and minor axes, $\lambda=c/a$, quantifying the eccentricity of an ellipsoid. Eq.(5) describes the translational motion of particles, where $\boldsymbol{x}$, $\boldsymbol{v}$, and $\boldsymbol{u}$ are the particle position, velocity, and fluid velocity at the particle location, respectively. Particles swim with a constant velocity $v_{\mathrm{swim}}$ relatively to the fluid in the direction of particle symmetric axis $\boldsymbol{p}$. The Stokes settling velocity $\boldsymbol{v}_{\mathrm{s}}$ is defined as follows [33, 34]:

$$\boldsymbol{v}_{\mathrm{s}} = v_{\mathrm{settle}}\left(\frac{\pi}{2}\right)\boldsymbol{e}_{\mathrm{g}} + \left[v_{\mathrm{settle}}(0) - v_{\mathrm{settle}}\left(\frac{\pi}{2}\right)\right](\boldsymbol{e}_{\mathrm{g}}\cdot\boldsymbol{p})\boldsymbol{p}\,, \tag{6}$$

where $\boldsymbol{e}_{\mathrm{g}}$ is the unit vector along the direction of gravity, and $v_{\mathrm{settle}}(\theta_{\mathrm{g}})$ is the Stokes settling velocity of an ellipsoid in a quiescent fluid with a fixed angle $\theta_{\mathrm{g}}$ between the particle major axis and the gravity direction. Here, $v_{\mathrm{settle}}$ also depends on the particle-to-fluid density ratio, particle size, and aspect ratio, and its expression was given by Siewert et al. (2014) [35]. As the Stokes settling velocity varies with particle shape, we change the particle size to ensure that particles settle with the same average velocity in the case when they have random orientation (see Table 2). The term $\boldsymbol{d}_{\mathrm{r}}$ in Eq.(4) and $\boldsymbol{d}_{\mathrm{t}}$ in Eq.(5) are Gaussian noises introduced to eliminate the effect of the particle initial condition on swimming trajectories. The noises are minor compared to hydrodynamic effects. The details of the numerical method are provided in the Appendix.

In the present study, the *one step Q-learning* algorithm is employed to iteratively approximate the optimal strategy [36]. A smart particle is considered as an agent, which is able to sense and receive information from the ambient flow (environment). The agent's state at the $n^{\mathrm{th}}$ state change ($s_n$) is determined by its instantaneous orientation and vorticity of the local flow. Then, the agent takes an action $a_n$ (i.e., selecting different preferential orientation $\boldsymbol{k}$) based on the current state according to a strategy. A Q-table, $Q(s,a)$, embodying the swimming strategy, stores the estimated values of every possible state-action pair. The agent takes the action with the highest estimated reward in most of the case, but it is also allowed to take other actions with small probability $\varepsilon$. The state space is the combination of four possible particle orientations (i.e. down, right, up, and left) and three possible vorticity levels (i.e. positive, negative, and approximately zero). Actions are taken among four preferential orientations [i.e. $\boldsymbol{k} = (k_x, k_y, 0)$, where $(k_x, k_y) = (0, -1), (1, 0), (0, 1)$, or $(-1, 0)$]. Configuration of the state and action is the same as in the study performed by Colabrese *et al.* (2017) [24], while detailed definitions are also provided in Appendix. At the next state change $s_{n+1}$, a new position of the agent is calculated by solving Eq. (4) and (5). An instant reward $r_n$ is then given by:

$$r_n = (y_{n+1} - y_n)/L_0\,, \tag{7}$$

where $y_n$ is the particle vertical coordinate when the $n$th state happens. This reward enables the algorithm to maximize the particle vertical velocity, and, thereby, the training converges to a strategy focused on the goal of swimming upwards quickly. Accordingly, the Q-table is updated using the following equation [37]:

$$Q(s_n, a_n) = Q(s_n, a_n) + \alpha\left[r_n + \gamma \max_a Q(s_{n+1}, a) - Q(s_n, a_n)\right]. \tag{8}$$

The learning rate $\alpha$ is chosen manually to control the convergence speed. The discount rate $\gamma$ is set at 0.999 in the present study allowing the algorithm to consider the most of the future rewards aiming to obtain a far-sighted strategy. To promote exploration of strategies, an optimistic initial value of $Q$ is set to be large relative to the converged value. The training process is divided into several episodes, and in each of them particles are initialized with the random location and orientation, and then, swim in the flow for $2\times10^4$ time steps. The strategy $Q$ is updated whenever the state of a particle is changed and converges to an approximately optimal strategy after a sufficient number of episodes. The detailed training parameters can be found in the Appendix. In each case, we conduct 50 independent trainings and choose the best one (with the highest vertical velocity) to analyze. Hereinafter, all the results and discussion are based on the best training.

In previous studies [24-26], only one particle is simulated in the training process. However, a large number of episodes is required for training a single particle, as states of a single particle change with a low frequency due to the relatively slow motion of particles. Therefore, we adopt multi-agent RL to train 50 randomly initialized particles simultaneously and allow them to share the same $Q$-table. The update frequency of $Q$ is enhanced, so that the convergence is accelerated. The performance of the learned strategy is not affected compared to training with a single particle under appropriate training parameters [38]. The flow field considered in the present study is simple and requires no computation cost. However, multi-agent learning might be critical in the case when the simulation of a complex fluid phase becomes a bottleneck in the training process.

## III. Results

In this section, we discuss the effect of the shape and gravity on particle swimming strategies in a TGV flow. Figure 1 shows the time-averaged particle velocity corresponding to different shapes, swimming strategies, and settling velocities. It can be seen that the RL strategies outperform the naive gyrotactic one in all cases, which implies that the naive gyrotactic strategy is not optimal in the present parameter space. It should be noted that in the case of the aspect ratio less than two, the RL strategies have the performance similar to that of the gyrotactic one regardless of settling. However, when the aspect ratio is greater than two, non-settling smart particles learn a smart strategy significantly better than the naive gyrotactic one. However, in the case of settling smart particles, smart particles only slightly outperform the naive ones. This result indicates the importance of both shape and gravity effects on a swimming strategy in a TGV flow.

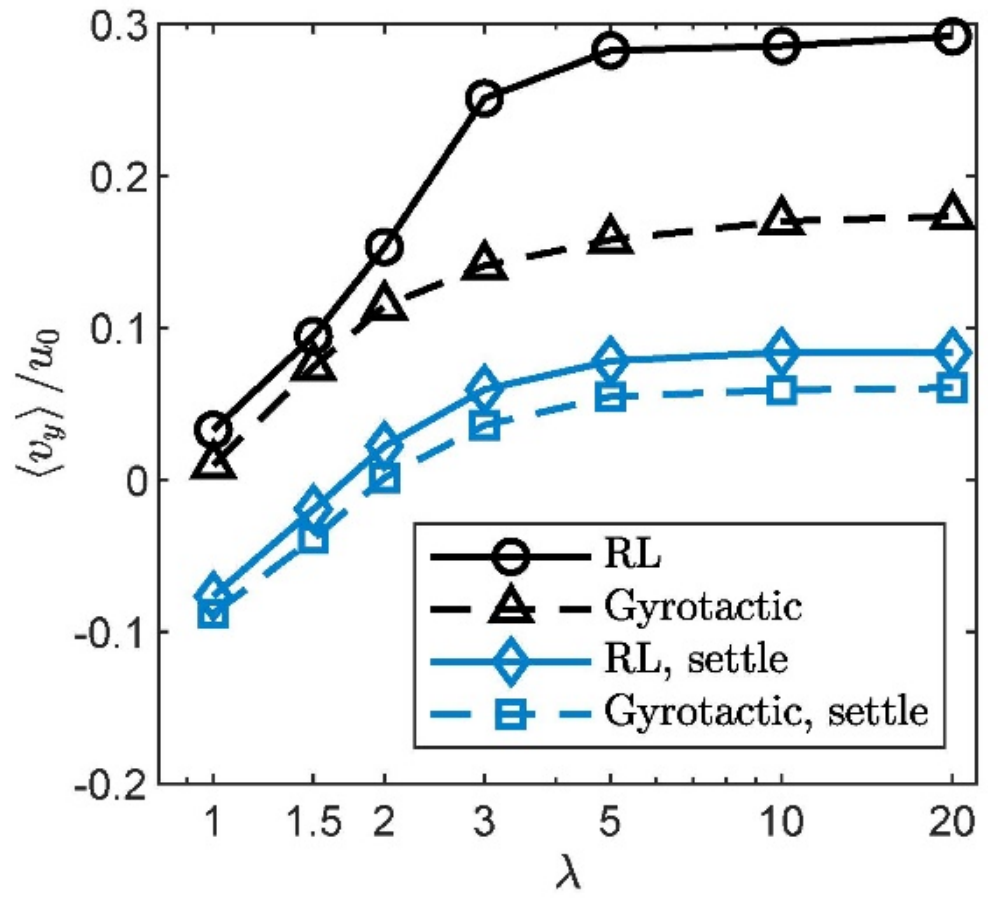

**Figure 1** Time-averaged vertical velocity of particles from $t$ = 0 to 50 s. RL (solid lines): RL strategy. Gyrotactic (dashed lines): Naive gyrotactic strategy. Circle and triangle (black): non-settling particles. Diamond and square (blue): settling particles.

## A. Shape effect

First, we neglect the settling velocity and consider particles with different aspect ratios. Figure 1 represents that the performance of RL and gyrotactic strategies increases as the particle aspect ratio rises and reaches a plateau after $\lambda = 5$. To obtain better understanding of why elongated particles swim upwards more efficiently, we analyze the distribution of particles, as presented in Figure 2. Both naive gyrotactic and smart particles of the spherical shape exhibit almost random distribution of the position and orientation in accordance with previous studies [7, 24]. As reorientation timescale $B$ is large, particle orientation is dominated by vorticity that disturbs the particle preferential orientation. In this case, both smart and gyrotactic particles are not able to swim upwards efficiently. However, as the aspect ratio increases, both smart and naive particles concentrate outside vortexes, and smart particles form clusters in an ascent path in the flow [Figure 2 (b, c)]. A quantitative measurement of clustering is given by a probability distribution function (PDF) in Figure 3. Both smart and naive particles remain almost random distribution when the aspect ratio is equal to one, However, they form a stronger cluster as the aspect ratio increases, which is indicated by a higher probability at low vorticity magnitude [Figure 3 (a, d)]. Clustering within the low-vorticity regions allows particles to avoid the disturbance in orientation caused by vortexes, which enables them to gradually align their orientation with the preferential alignment. Moreover, smart particles learn to sample upwelling regions, where the background flow carries them upwards [Figure 3 (b)]. Both mechanisms promote an elongated particle to swim upwards, suggesting that the RL approach captures the features of a TGV flow appropriately, and develops a strategy to take advantage of these features.

To analyze the swimming strategy, in Figure 4, we show a typical trajectory and the Q-table of a particle of $\lambda = 5$, whose shape effect saturates. The particle initialized in an anticlockwise vortex keeps aligning rightwards. Although the weak active alignment cannot overcome the vorticity effect, the particle succeeds to swim out of vortex and reaches an ascent path where the local flow goes upwards. After this, the particle seeks to align upwards and rightwards alternately. Interestingly, particles tend to align with a direction that is orthogonal to the local flow direction. However, orientation is dominated by the fluid deformation rate due to the low vorticity and the small contribution of active alignment. When a particle approaches the stagnation point, it is forced to rotate to an orthogonal direction under the strain rate effect. In this procedure, both strain rate effect and active alignment play important roles. As particles tend to stay outside vortexes, the orientation is dominated by the strain rate term in Eq.(4), which can be also written as follows:

$$\frac{\lambda^2-1}{\lambda^2+1}[\boldsymbol{I}-\boldsymbol{pp}]\cdot\boldsymbol{S}\cdot\boldsymbol{p}=\frac{\lambda^2-1}{\lambda^2+1}\boldsymbol{\omega}_s\times\boldsymbol{p}, \tag{9a}$$

where $\boldsymbol{\omega}_s=\boldsymbol{p}\times(\boldsymbol{S}\cdot\boldsymbol{p})$. (9b)

Here, $\boldsymbol{\omega}_s$ denotes the particle angular velocity resulting from the strain rate, which aligns $\boldsymbol{p}$ with $\boldsymbol{S}\cdot\boldsymbol{p}$. In the TGV flow, shear components of the strain rate tensor are zero, and normal components obey $S_{11}=-S_{22}$ and $S_{33}=0$. This relationship makes the directions of $\boldsymbol{S}\cdot\boldsymbol{p}$ and $\boldsymbol{p}$ symmetric with $x$- or $y$-axis depending on the sign of $S_{11}$. Therefore, the strain rate effect in TGV aligns particle with directions of coordinate axes, which explains the zigzag clustering pattern observed in Figure 2 (c). The alignment due to the strain rate stabilizes particle orientation against the vorticity disturbance and allows particle to gain a better control of the swimming direction. Therefore, active alignment of smart particles is also important. Although the active term in Eq.(4) is small relative to the strain rate term, its accumulated contribution determines which direction a particle will orientate to when it approaches a stagnant point. Smart particles choose the active alignment that is nearly perpendicular to instantaneous orientation to maximize the contribution of the active term. As a result, smart particles can adjust their orientation to stay in the ascent path of the flow to move upwards. It should be noted that smart particles do not always learn to form "right-up" clusters. Due to the symmetry of left and right in a TGV, particles can also learn a "left-up" strategy with the same probability. However, we do not show the results of "left-up" strategies in the present paper for conciseness.

In short, particle orientation is dominated by the fluid strain rate, and the direction of transportation is decided by active choices of a particle. This explains the increasing performance of smart particles when the aspect ratio increases. A larger aspect ratio results in better stability, making it easier for particles to entry low-vorticity regions. In such regions, smart particles can adjust their swimming direction and stay in the ascent path. However, the shape effect saturates when $\lambda > 5$, as the coefficient of the strain rate term, $(\lambda^2-1)/(\lambda^2+1)$, also saturates with increasing $\lambda$. This explains the plateau of performance (see Figure 1).

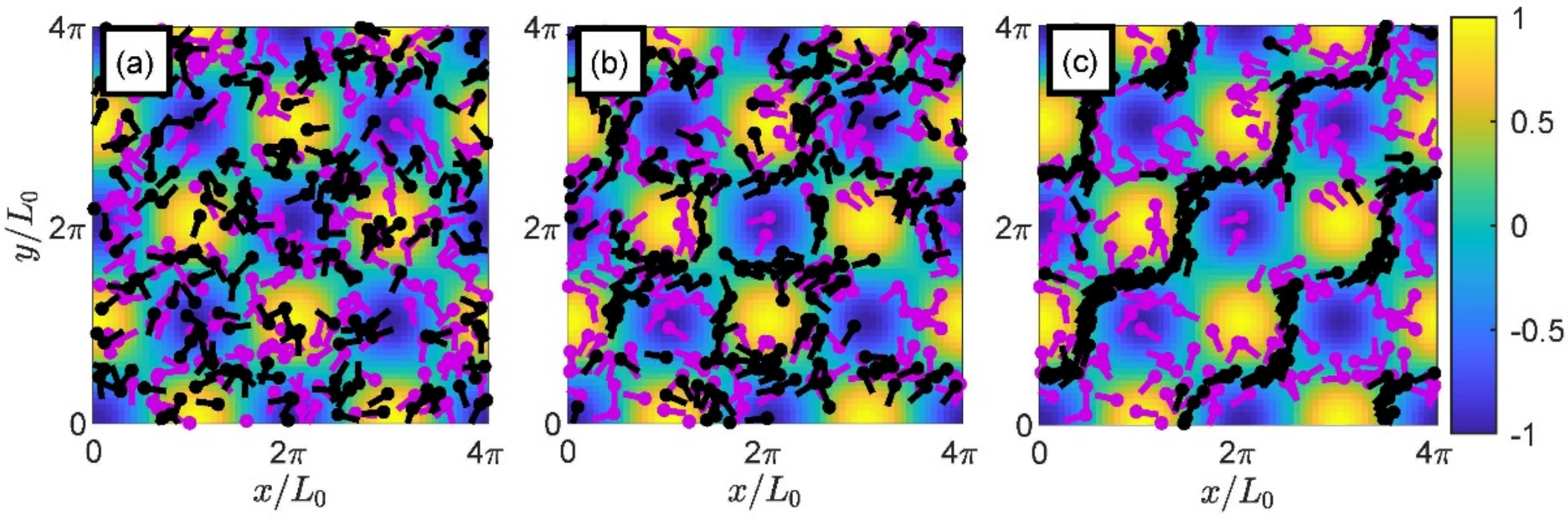

**Figure 2** Instantaneous distribution of particles swimming with RL (black) and naive gyrotactic (purple) strategies at t = 50 s. (a) $\lambda = 1$, (b) $\lambda = 2$, (c) $\lambda = 5$. Settling velocity is neglected. Dots indicate the positions of particles, while segments denote the "tails" that pointing to the direction opposite to the swimming. The background contour is colored according to the flow vorticity.

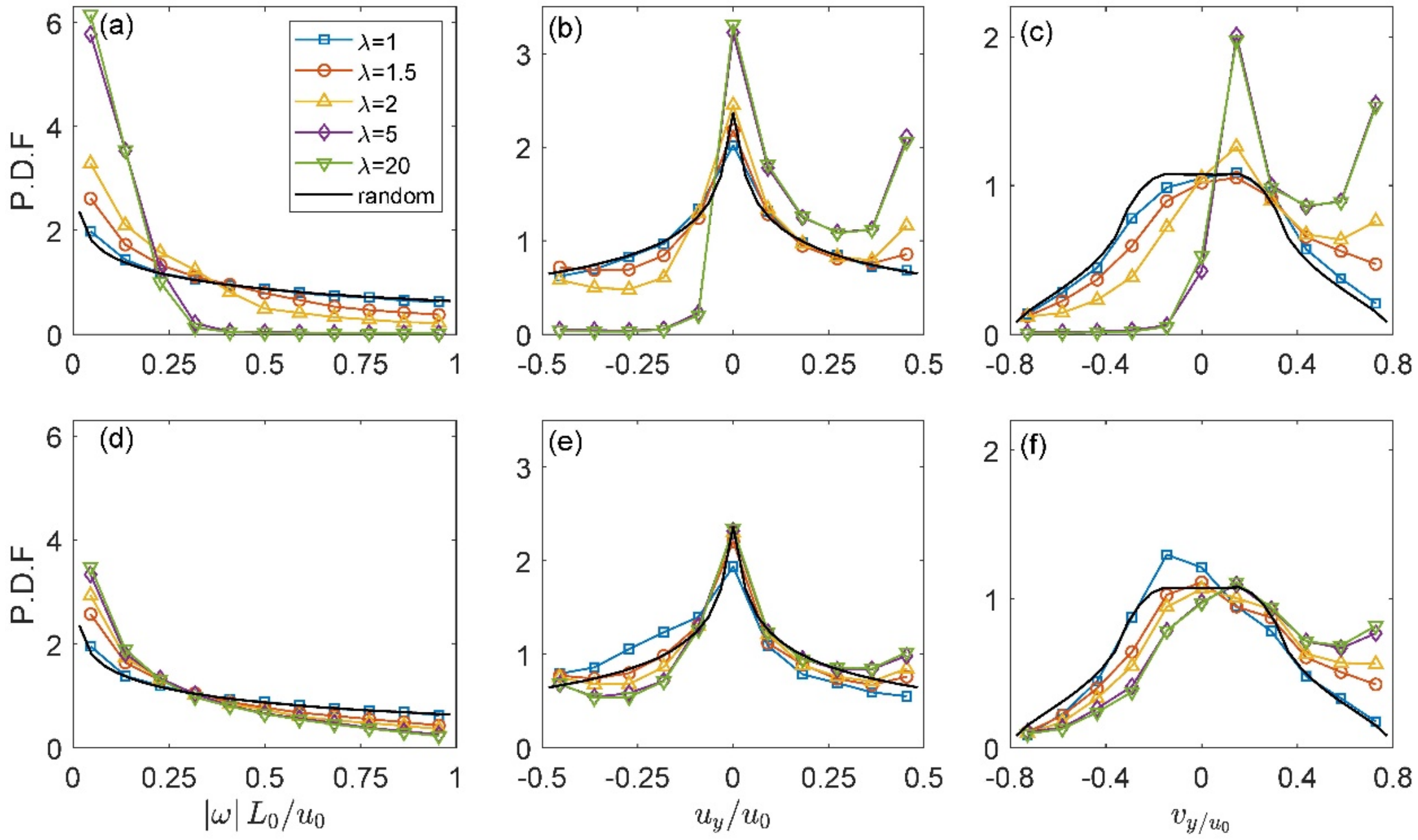

**Figure 3** Probability distribution function (t = 50 s) of (a, d) absolute vorticity, (b, e) vertical flow velocity at particle locations, and (c, f) vertical particle velocity. (a, b, c): RL strategy. (d, e, f) Naive strategy.

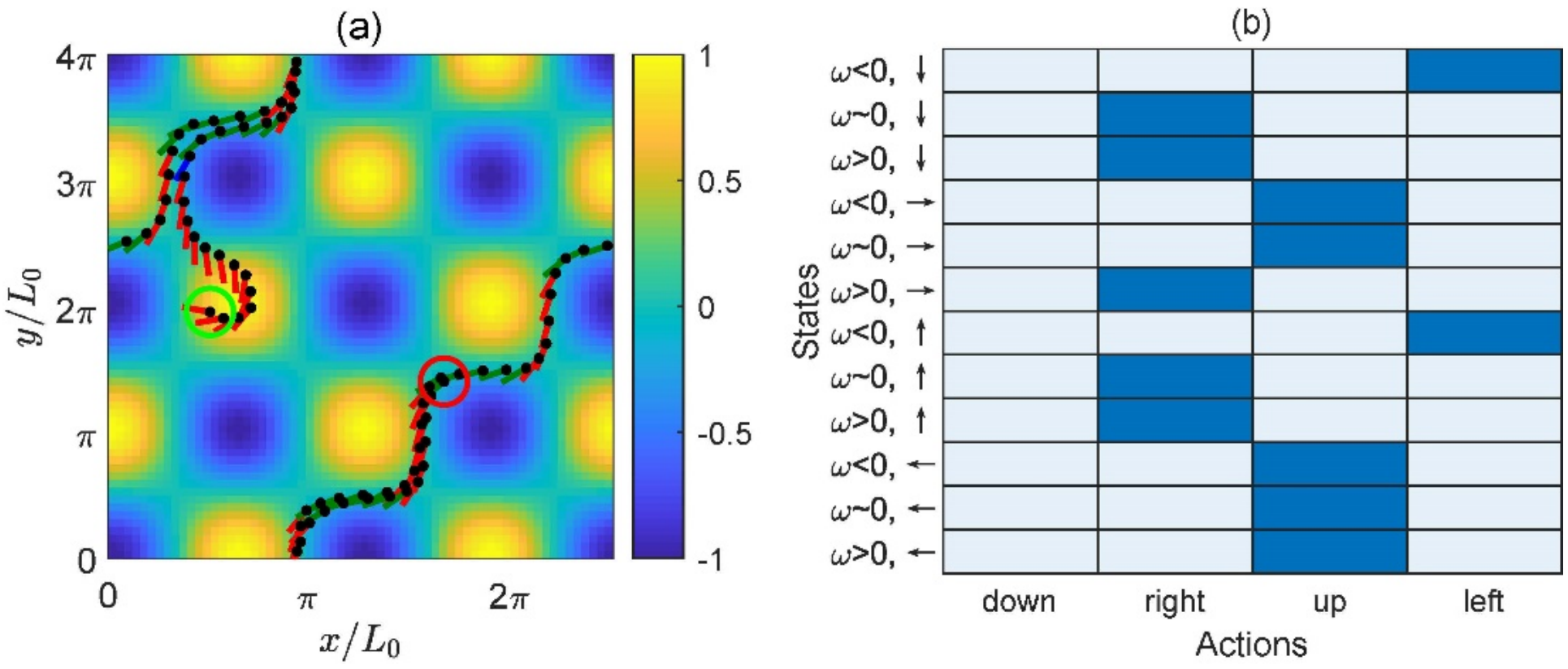

**Figure 4** A non-settling elongated particle (λ = 5). (a) A typical trajectory from $t = 0$ to 15s. Black dots denote the position of particle. Segments denote the "tails" pointing to the direction opposite to the swimming direction. Segments are colored according to particle instantaneous actions of aligning down (not appear in this trajectory), right (red), up (green), and left (blue). Green and red circles indicate the start and the end of trajectory, respectively. (b) Q-table of the RL swimming strategy. Rows: different states, where $\omega$ and arrows indicate three vorticity levels and four particle orientations, respectively. Columns: actions taken in each state are painted in blue.

## B. Gravity effect

When gravity is exerted on particles, the performance of both RL and naive gyrotactic strategies deteriorates because of the settling velocity. However, the trend is not altered (Figure 1). When the aspect ratio is greater than two, smart particles start to learn strategies to swim upwards against gravity owing to the strain rate effect. However, the advantage of actively choosing a swimming direction does not lead to a smart particle outperforming naive gyrotactic ones significantly, even when the aspect ratio is greater than five. This indicates that the learned strategies might be similar to the gyrotactic one when particles subject to settling. In order to discuss the gravity effect, we focus on the particle of λ = 5, for which the shape effect has been saturated for both RL and gyrotactic strategies.

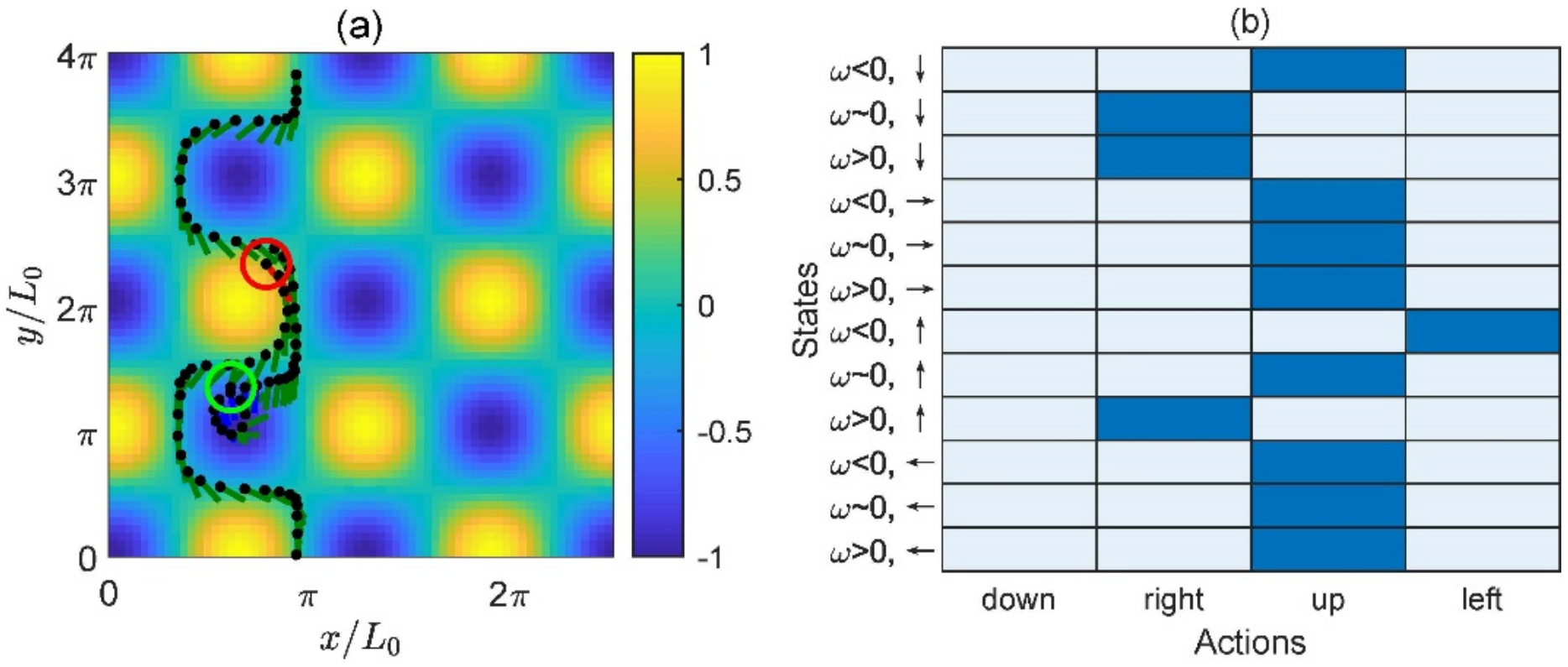

**Figure 5** A settling elongated particle (λ = 5). (a) Typical trajectory from $t$ = 0 to 15 s. Black dots denote the position of the particle. Segments denote the "tails" pointing to the direction opposite to the swimming direction. Segments are colored according to particle instantaneous actions of aligning down (not appear in this trajectory), right (red), up (green), and left (blue). Green and red circles indicate the start and the end of trajectory, respectively. (b) Q-table of the RL swimming strategy. Rows: different states, where $\omega$ and arrows indicate three vorticity levels and four particle orientation, respectively. Columns: actions taken in each state are painted in blue.

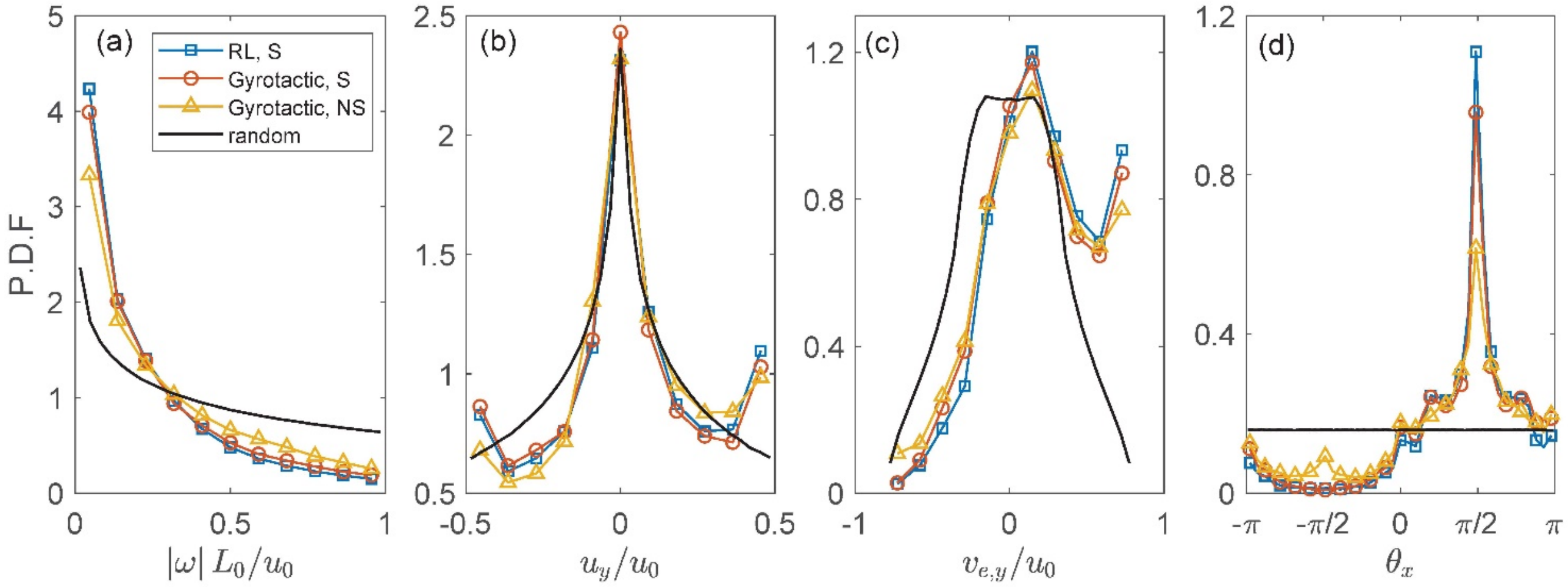

**Figure 6** Probability distribution functions of particles with λ = 5 at t = 50 s. RL: RL strategy. Gyrotactic: Gyrotactic strategy. S: Settling particles. NS: Non-settling particles. (a) Absolute value of vorticity at particle locations. (b) Vertical flow velocity at particle locations. (c) Vertical component of the settling-excluded particle velocity. For settling particles, the vertical settling-excluded velocity is greater than the particle velocity, $v_y$, as the settling velocity is negative. For non-settling particles, the two velocities are equivalent. (d) Angle between particle orientation and $x$-axis.

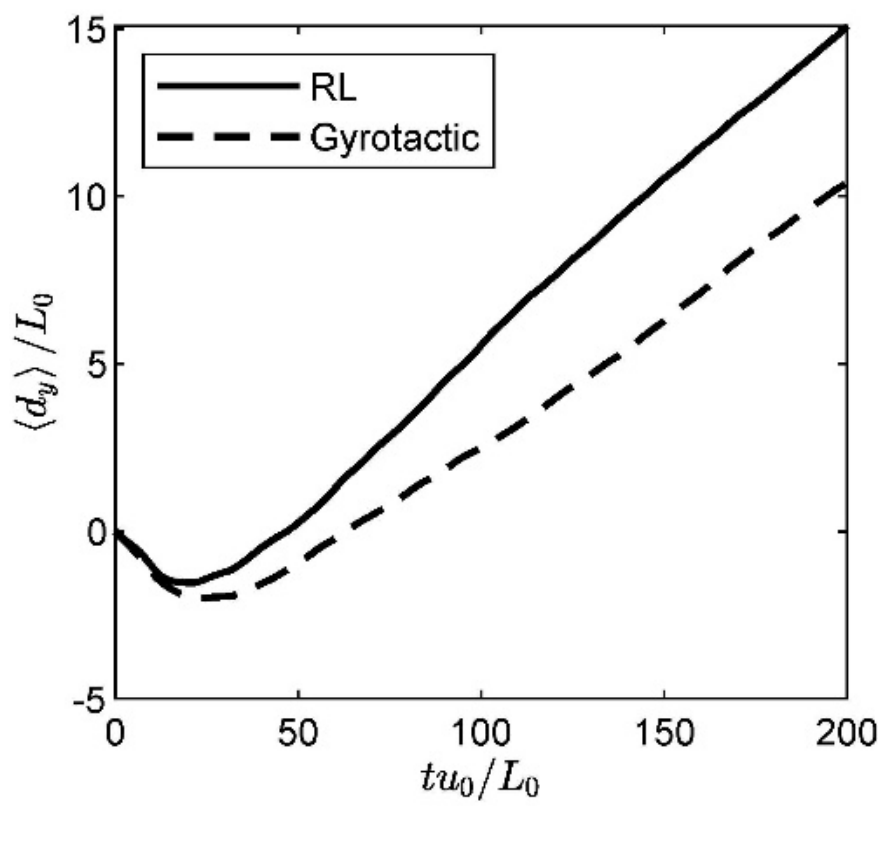

**Figure 7** Mean particle vertical displacement $<d_y>$ over the time. RL and Gyrotactic denote RL and naive gyrotactic strategies, respectively.

In Figure 5, we show a trajectory and the Q-table of the RL strategy corresponding to such a settling elongated particle. A smart particle tries to align upwards in most of states, unless it orientates downwards, or already orientates upwards but locate in high-vorticity regions. Due to the large reorientation timescale, a particle cannot perform rapid changes in orientation. Therefore, the smart particle learns to align upwards more often, because it needs to swimming against gravity. Settling smart particles can no longer form clusters as non-settling particles do, as a settling particle probably drops in vortexes if they align horizontally. This RL strategy results in a similar particle distribution as that of gyrotactic particles, as both kind of particles have the same preferential orientation to the upward direction in most of the cases. As shown in Figure 6 (a, d), most of smart particles locate in low-vorticity regions and align directly upwards. Therefore, there is no considerable difference in PDF of the position and orientation between smart and naive gyrotactic particles after they obtain steady orientation. In spite of the similarity, the RL strategy performs better, as shown by the mean vertical displacement over time in Figure 7. Aligning rightwards while orientating downwards enables smart particles to obtain an upward orientation more quickly, and thereby, smart particles start to move upwards earlier than gyrotactic ones as shown in Figure 7. The RL strategy also allows particles to stay in low-vorticity regions and maintain their upward orientation, which explains a higher peak of PDF in Figure 6 (d) and a higher slope of the mean vertical position in Figure 7 than those of naive gyrotactic particles.

The similarity between the RL and naive gyrotactic strategies for settling particles is an interesting observation. As the RL algorithm converges to the approximately optimal strategy for a given task, the obtained results suggest that gyrotaxis might be considered as a simple but effective strategy for a settling, elongated particle to swim upwards. Non-settling smart particles can learn a smart strategy that is much better than gyrotactic one. Nevertheless, with inclusion of settling velocity, the performance of smart strategy becomes similar to that of gyrotactic one as discussed in sect. 3.1. This contrast means that the gyrotaxis is basically far away from the optimal strategy in the case without gravity, but it is close to the optimal strategy when we consider settling velocity. Therefore, we suggest that the gyrotactic strategy is more suitable for settling particles rather than for non-settling ones. To verify this hypothesis, distributions of gyrotactic particles in settling and non-settling cases are compared in Figure 6 (Gyrotactic, S and Gyrotactic, NS). We define the settling-excluded particle velocity, $v_{e,y} = (\boldsymbol{v}\text{-}\boldsymbol{v}_s)\cdot\boldsymbol{e}_y$, to compare particles with different settling velocities [Figure 6 (c)]. It is observed that settling gyrotactic particles have a greater vertical settling-excluded velocity than non-settling gyrotactic particles, which means the former employ their gyrotaxis and swimming velocity to obtain an upward velocity in a better manner. Although they show weaker sampling of upwelling regions, they form stronger clustering in low-vorticity regions [Figure 6 (a)] and stronger preferential alignment to the upward direction [Figure 6 (d)].

## IV. Discussion and conclusion

In the present study, we adopt the Q-learning algorithm to obtain approximately optimal swimming strategies to swim upwards in a TGV flow for inertia-less particles of different shapes considering the effect of the settling velocity caused by gravity. Investigation conducted on aspherical microswimmers reveals that the shape and gravity effects, which are common in reality, are important to the swimming strategy of microswimmers in the TGV flow. Shape effect is significant for particles with large reorientation timescale.

While spherical particles have almost random distribution and the behavior similar to that of naive gyrotactic particles, elongated particles learn to accumulate in a low-vorticity region and take advantage of the background flow to move upwards efficiently. Elongated particles are more stable owing to the deformation rate, which reduces the perturbation of vorticity on their orientation. Both RL and naive gyrotactic strategies show better performance with an increase in the aspect ratio. However, the performance reaches a plateau as the coefficient of the deformation term in Eq. (4) saturates. When gravity is taken into consideration, the performance of both smart and gyrotactic particles declines due to the influence of the settling velocity. While spherical particles keep sinking due to the gravity, elongated particles can learn a strategy to swim with a positive vertical velocity against the settling effect. However, the learned strategies are similar to the gyrotactic one. In addition, we show that gyrotactic is more suitable for a settling particle than for a non-settling one. Although the gyrotactic strategy is inefficient for a non-settling particle, it performs almost as good as the RL strategy when particles are settling.

Two interesting observations are presented in the obtained results. First, it is seen that non-settling particles with the aspect ratio less than two perform similarly to gyrotactic ones. Second, settling particles follow the strategies that are similar to those of the gyrotactic ones. These observations indicate that gyrotaxis might be considered as a simple but effective strategy for slightly spheroidal settling particles to swim vertically in a simple steady TGV flow. As the considered parameters of particles and the flow are defined based on realistic marine plankton properties and the marine turbulence flow scales, it is logical to suppose that gyrotaxis is also an appropriate strategy for realistic, slightly aspherical plankton to move vertically in an vortical aquatic environment. This leads to a conjecture that gyrotactic plankton, subject to the gravity and shape effects, have "learned" the gyrotactic strategy by evolving the bottom-heavy mass distribution to obtain gyrotaxis as a result of the natural selection. However, in the real marine environment, the temporal and spatial fluctuations of turbulence may considerably affect the behavior of gyrotactic particles [12, 39]. In this case, gyrotaxis might be suboptimal, and plankton may develop a certain reaction to the turbulent environment. Moreover, the results presented in Figure 6 also show that gyrotaxis is deemed more important to settling particles than non-settling ones. Considering the fact that many of marine plankton have considerable settling velocity [28], gyrotaxis may have greater significance to the vertical migration of plankton than we expected. However, the coupling of settling velocity and gyrotaxis for microswimmers in a turbulent environment still remains unknown. Therefore, the corresponding swimming strategy and exact mechanisms of gyrotaxis of marine microorganisms still require further investigation.

## Acknowledgement

This work was supported by the National Natural Science Foundation of China (Grant Nos. 11911530141, 11772172, and 91752205). JQ, WH, and LZ acknowledge the support from the Institute for Guo Qiang of Tsinghua University (project No. 2019GQG1012).

## Appendix

In the conducted simulation, the governing equations of particle orientation and position, Eqs. (4) and (5), are solved by the second-order Adam-Bashforth scheme. The time step size is $\Delta t=0.01L_0/u_0$. The noise terms $\boldsymbol{d}_\mathrm{r}$ and $\boldsymbol{d}_\mathrm{t}$ are used to eliminate the effect of the particle initial condition, because the flow is steady. Noise term of orientation is added on the orientation solved by scheme, $\tilde{\boldsymbol{p}}_n$, at every time step:

$$\boldsymbol{p}_n = \tilde{\boldsymbol{p}}_n + \boldsymbol{d}_\mathrm{r}, \tag{a1}$$

where $\boldsymbol{p}_\mathrm{n}$ denotes the orientation at $n^\mathrm{th}$ time step. The term $\boldsymbol{d}_\mathrm{r}$ is equivalent to rotating the particle orientation for a tiny angle, which is calculated as follows:

$$\boldsymbol{d}_\mathrm{r} = \left(-\theta_\mathrm{r}\tilde{p}_{n,x}, \theta_\mathrm{r}\tilde{p}_{n,y}, 0\right)^\mathrm{T}, \tag{a2}$$

where $\theta_\mathrm{r} = N\sqrt{2D_\mathrm{r}\Delta t}$. (a3)

Here, $N$ is an independent standard normal random variable, $D_\mathrm{r} = 4\times10^{-4}$ s$^{-1}$ is the rotational diffusion constant, and $\Delta t$ is the time step size. In this way, the strength of diffusion is not correlated with the time step size. Similarly, translational noise terms producing a small random displacement are calculated as follows:

$$\boldsymbol{d}_\mathrm{t} = N'\sqrt{2D_\mathrm{t}\Delta t}\left(\cos\theta_\mathrm{d}, \sin\theta_\mathrm{d}, 0\right)^\mathrm{T}, \tag{a4}$$

where $N'$ is an independent standard normal random variable, $D_\mathrm{t} = 1\times10^{-9}$ m$^2$/s is the translational diffusion constant, and $\theta_\mathrm{d}$ is a uniform random variable to determine the direction of translational diffusion.

In Q-learning, states of particles are determined by the instantaneous orientation and local vorticity. We have twelve states in total. Orientation falls into four substates as follows:

$$s_\mathrm{o}\left(\theta_x\right) = \begin{cases} 1, & \theta_x \in \left[5\pi/4, 7\pi/4\right), \\ 2, & \theta_x \in \left[0, \pi/4\right) \cup \left[7\pi/4, 2\pi\right), \\ 3, & \theta_x \in \left[\pi/4, 3\pi/4\right), \\ 4, & \theta_x \in \left[3\pi/4, 5\pi/4\right), \end{cases} \tag{a5}$$

where $s_\mathrm{o}$=1, 2, 3, or 4 represents four orientation states (i.e. down, right, up, or left), respectively. Local vorticity is divided into three levels:

$$s_\mathrm{v} = \begin{cases} 1, & \omega_z L_0/u_0 < -1/3, \\ 2, & -1/3 \le \omega_z L_0/u_0 \le 1/3, \\ 3, & \omega_z L_0/u_0 > 1/3, \end{cases} \tag{a6}$$

where $s_\mathrm{v}$=1, 2, or 3 represents negative, approximately zero, or positive vorticity, respectively. Learning rate, $\alpha$, and exploration rate (probability to take a suboptimal action), $\varepsilon$, decrease according to the episode number, $E$, i.e.:

$$\alpha = \frac{\alpha_0}{1 + E/\sigma_0}, \tag{a7}$$

$$\varepsilon = \varepsilon_0 \max\left(0, \frac{E_0 - E}{E_0}\right), \tag{a8}$$

where $\alpha_0$ and $\varepsilon_0$ are the initial values of the learning rate and exploration rate, respectively. Here, $\sigma_0$ and $E_0$ are constants to control the decreasing rate. The initial value of the Q-table is 2000.0. The training parameters of each case are given in Table a1.

**Table a1** Training parameters.

| Case name | $\lambda$ | settling | $\alpha_0$ | $\varepsilon_0$ | $\sigma_0$ | $E_0$ |
|---|---|---|---|---|---|---|
| cnl1 | 1.0 | No | 0.20 | 0.005 | 400 | 800 |
| cnlp5 | 1.5 | No | 0.12 | 0.005 | 400 | 800 |
| cnl2 | 2.0 | No | 0.12 | 0.005 | 400 | 800 |
| cnl3 | 3.0 | No | 0.12 | 0.005 | 400 | 800 |
| cnl5 | 5.0 | No | 0.10 | 0.005 | 400 | 800 |
| cnl10 | 10.0 | No | 0.10 | 0.005 | 400 | 800 |
| cnl20 | 20.0 | No | 0.05 | 0.005 | 400 | 800 |
| csl1 | 1.0 | Yes | 0.20 | 0.005 | 400 | 800 |
| cslp5 | 1.5 | Yes | 0.18 | 0.005 | 400 | 800 |
| csl2 | 2.0 | Yes | 0.12 | 0.005 | 400 | 800 |
| csl3 | 3.0 | Yes | 0.12 | 0.005 | 400 | 800 |
| csl5 | 5.0 | Yes | 0.10 | 0.005 | 400 | 800 |
| csl10 | 10.0 | Yes | 0.10 | 0.005 | 400 | 800 |
| csl20 | 20.0 | Yes | 0.05 | 0.005 | 400 | 800 |